\documentclass[twocolumn,showpacs,preprintnumbers,amsmath,amssymb]{revtex4}

\pdfoutput=1
\usepackage{xspace}
\usepackage{epsfig}
\usepackage{amsfonts}
\usepackage{graphicx}
\usepackage{url}
\usepackage[dvipdfm,%
              colorlinks,%
          linkcolor=blue,%
        citecolor = blue,%
              hyperindex,%
        plainpages=false,%
         urlcolor = blue,%
       pdfstartview=FitH]{hyperref}

\begin{document}


\title{\textsf{High-field Current-carrying Capacity of Semiconducting Carbon Nanotubes}}

\author{Debdeep Jena}

\affiliation{Department of Electrical Engineering, University of Notre Dame, IN, 46556, USA}

\date{\today}

\begin{abstract}
It is shown that current saturation in semiconducting carbon nanotubes is indistinguishable from metallic nanotubes if the carrier density is above a critical value determined by the bandgap and the LO phonon energy.  This feature stems from the higher number of current-carrying states in the semiconducting tubes due to the van-Hove singularity at the band-edge.  Above this critical carrier density, the ensemble saturation velocity at high-fields is found to be independent of the bandgap, but strongly dependent on the carrier density, explaining recent observations.
\end{abstract}



\maketitle


{\em Introduction:} Carbon nanotubes (CNTs) exhibit exceptional current carrying capability.  For example, Yao et al. in a seminal work on single-wall metallic CNTs (m-CNT) observed that the current carrying capacity, as high as $10^{9}$ A/cm$^2$, was fundamentally limited by the emission of LO phonons by energetic electrons at high electric fields \cite{prl00dekkerCNTsatCurrent}.  The exceptionally high current carrying capacity was attributed by the authors to the fact that each carrier within a energy bandwidth determined by the LO phonon energy $\hbar \omega_{LO} \sim 160$ meV moves with the Fermi velocity ($v_{F} \sim 10^{8}$ cm/s), leading to a fundamental limit of the current determined by $I_{0} =( 4 e/L )\sum_{k} f_{k} v_{F} = 4 e f_{LO}$, where $f_{LO} = \omega_{LO}/2 \pi$, $L$ is the CNT length, the factor $g = g_{s} g_{v} = 4$ is the product of the spin and valley degeneracy, and $f_{k}$ is the occupation probability of state $| k \rangle $.  The current evaluates to $\approx 25$ $\mu$A, precisely explaining the experimental saturation current.  This result is fundamental for long m-CNTs, where transport is in the diffusive limit; for short m-CNTs, the current can exceed the diffusive limit of 25 $\mu$A due to ultrafast phonon emission leading to non-uniform heating.  It has been shown that the increase in the saturation current over this limit can be explained by consideration of non-equilibrium (hot) phonons  \cite{prl05leburtonCNTsatCurrent, prb06lazzeriCNTsatCurrent}.  

There have been reports of similar current saturation in semiconducting CNTs (s-CNT) as well \cite{nature03javeyCNTsatCurrent, prl04javeyCNTsatCurrent, prl05fuhrerCNTsatCurrent}.  However, the physical mechanism of current saturation is not yet clear.  Previous theoretical considerations indicate a velocity  saturation \cite{prl05avourisCNTsatCurrent}, and in \cite{pss06fuhrerCNTsatCurrent}, a dependence of the ensemble saturation velocity on the carrier density was established by the solution of the Boltzmann Transport Equation (BTE).  The purpose of this letter is fourfold: a) to show that for 1D s-CNTs {\em current saturation} is as fundamental as in m-CNTs, b) as opposed to m-CNTs, the saturation current in a s-CNT is sensitive to the location of the Fermi level $\mathcal{E}_{F}^{0}$ {\em before the application of a bias}, c) the net ensemble velocity (`saturation velocity') of carriers in s-CNTs is {\em not} universal, but strongly dependent on $\mathcal{E}_{F}^{0}$ as well, and can range from $0.1 - 1 \times v_{F}$, and d) the saturation velocity of carriers in s-CNTs becomes independent of the bandgap for large carrier densities.  These four facts are proved using analytical arguments based on the bandstructure, and the ultrafast electron-optical phonon interactions in CNTs.  They offer a simple explanation for various recent observations made from experiments and numerical simulations \cite{prl05avourisCNTsatCurrent}.

{\em CNT bandstructure \& saturation current:} For the analysis of transport properties, it suffices to consider expansions of the energy dispersion for small momenta around the $\mathcal{K}, \mathcal{K}^{\prime}$ points of the underlying graphene bandstructure \cite{rmp07charlierCNTtransport}.  Let the bandstructure of the $n^{th}$ subband of a s-CNT be given by the dispersion $\mathcal{E}_{n}(k_{x}) = \hbar v_{F} \sqrt{k_{x}^{2} + k_{n}^{2}}$, where $\hbar$ is the reduced Planck's constant, $k_{x}$ is the longitudinal wavevector, and $k_{n}$ is the transverse wavevector quantized by the diameter of the CNT.  All energies are measured with respect to the Dirac point of the underlying graphene bandstructure.  The bandgap of the CNT with this bandstructure is given by $\mathcal{E}_{g} = 2 \hbar v_{F} k_{0}$, where $k_{0}$ is the smallest allowed (nonzero) transverse wavevector, fixed by the size quantization ($\mathcal{E}_{g} \sim 800/d$ meV, where $d$ is the diameter of the s-CNT in nm).    

\begin{figure}
\begin{center}
\leavevmode \epsfxsize=3.3in \epsfbox{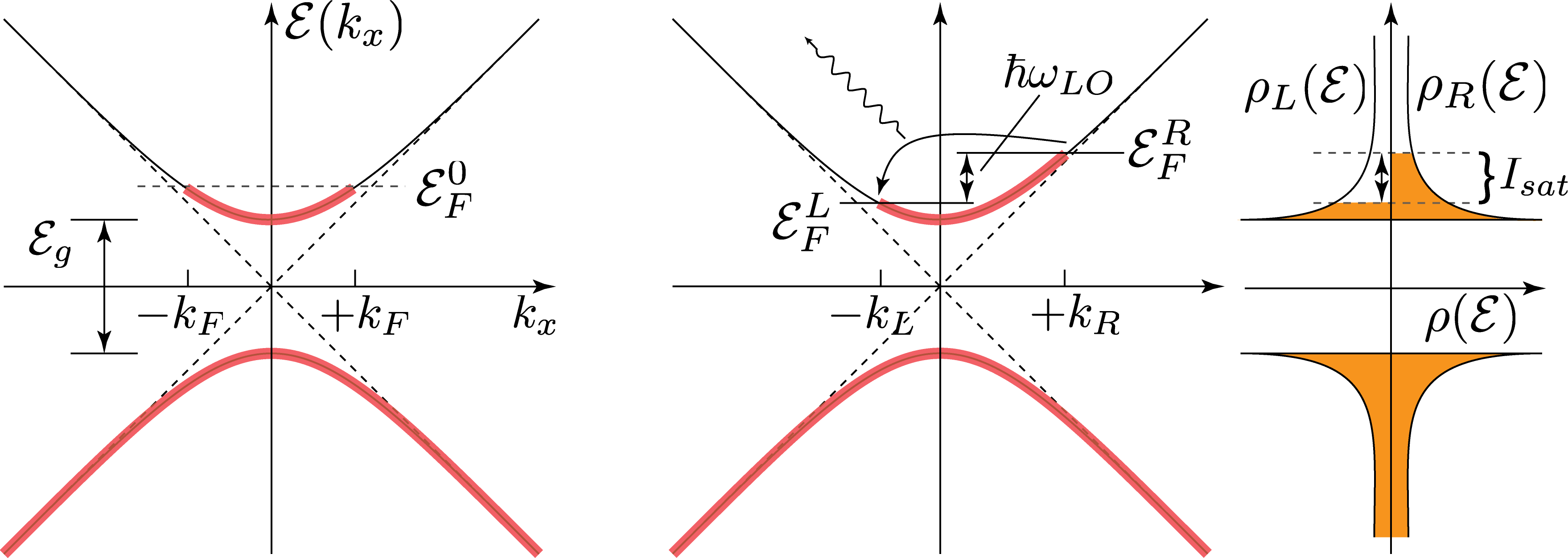}
\caption{Schematic representation of the mechanism of LO phonon scattering in s-CNTs.  If the Fermi level before the application of a bias is high in the conduction band, the ultrafast optical phonon scattering leads to the steady state carrier distribution shown in the middle and right figures; the saturation current is effectively carried by right-moving electrons spread over an energy bandwidth of $\hbar \omega_{LO}$.} 
\label{sCNT_kspace_hidensity}
\end{center}
\end{figure}

An electric field ${\bf F} = -F_{0} \hat{x}$ is applied across the drain-source contacts to the s-CNT, which are assumed to be perfectly transparent to electron flow.  To determine the exact shape of the new Fermi surface, one needs to solve the BTE, as has been done in \cite{pss06fuhrerCNTsatCurrent} for $T= 300$ K.  However, for the $T \rightarrow 0$ limit, various analytical results may be derived to illustrate the physics of current saturation in s-CNTs.  The Fermi surface at low temperatures is sharp, and the force due to the electric field $(-e) {\bf F} = \hbar d{\bf k}/dt$ populates $+k_{x}$ states by emptying out $-k_{x}$ states near the Fermi surface.  If the contacts do not inject carriers into the wire (no space-charge currents), the equilibrium occupation function $f_{k}^{0} = \Theta( k_{F} -  | k_{x} | )$ (where $\Theta (...)$ is the Heaviside unit-step function) shifts to the non-equilibrium position $f_{k} = 1$ for $-k_{L} \leq k_{x} \leq +k_{R}$, while conserving the particle number inside the wire.  Here $k_{L}, k_{R}$ are the wavevectors of the highest right-going and the highest left-going electrons in response to the electric field.  This is shown schematically in Fig \ref{sCNT_kspace_hidensity}.  

The density of states can be split equally into the left-going and right-going states: $\rho_{tot}(\mathcal{E}) = \rho_{L}(\mathcal{E}) + \rho_{R}(\mathcal{E})$, where 
\begin{equation}
\rho_{R,L}( \mathcal{E} )= \frac{1}{2} \cdot \frac{ g }{ \pi \hbar v_{F} } \cdot \frac{ \mathcal{E} }{ \sqrt{ \mathcal{E}^{2} - (\frac{ \mathcal{E}_{g} }{ 2 })^{2} } },
\end{equation}
and the resulting right- and left-going charge carrier densities can be then written in the form $n_{R,L} = \int_{\mathcal{E}_{g}/2}^{\mathcal{E}_{F}^{R,L}} \rho_{R,L}(\mathcal{E}) d\mathcal{E}$ as
\begin{equation}
n_{R,L} = \frac{1}{2} \cdot \frac{ g }{ \pi \hbar v_{F} } \cdot \sqrt{ (\mathcal{E}_{F}^{R,L})^{2} - (\frac{ \mathcal{E}_{g} }{ 2 })^{2} },
\end{equation}
which defines the quasi-Fermi levels $\mathcal{E}_{F}^{R}$ and $\mathcal{E}_{F}^{L}$.  Before the application of a bias, $\mathcal{E}_{F}^{R} = \mathcal{E}_{F}^{L} = \mathcal{E}_{F}^{0}$, and the 1D carrier density is given by
\begin{equation}
n_{0} = n_{R} + n_{L} = \frac{ g }{ \pi \hbar v_{F} } \cdot \sqrt{ (\mathcal{E}_{F}^{0})^{2} - (\frac{ \mathcal{E}_{g} }{ 2 })^{2} }.
\end{equation}
If the contacts to the s-CNT do not inject excess carriers into the tube, then to ensure particle number conservation $n_{0} = n_{R} + n_{L}$ requires that at all bias conditions, the relation
\begin{equation}
2 \sqrt{ (\mathcal{E}_{F}^{0})^{2} - (\frac{ \mathcal{E}_{g} }{ 2 })^{2} } = \sqrt{ (\mathcal{E}_{F}^{R})^{2} - (\frac{ \mathcal{E}_{g} }{ 2 })^{2} } + \sqrt{ (\mathcal{E}_{F}^{L})^{2} - (\frac{ \mathcal{E}_{g} }{ 2 })^{2} }
\label{chargebalance}
\end{equation}
must be satisfied for the quasi-Fermi levels.

At a high field, the distribution function reaches a stage when the difference between the highest filled electronic state (HFES) and the lowest empty electronic state (LEES) equals the LO phonon energy to allow for energy relaxation by the emission of optical phonons.  Building upon the theory in \cite{prl00dekkerCNTsatCurrent}, we make a {\em hypothesis} that if the LO phonon emission process is ultrafast, then the distribution function is locked in this configuration, and is resistant to any further increase in applied bias.  Depending on the availability of charge carriers, the LEES may be $\mathcal{E}_{F}^{L}$, or the band edge ($\mathcal{E} = \mathcal{E}_{g}/2$).  Before we proceed to calculate the net saturation current, we introduce some critical parameters that serve to highlight this fact. 

\begin{figure}
\begin{center}
\leavevmode \epsfxsize=2.7in \epsfbox{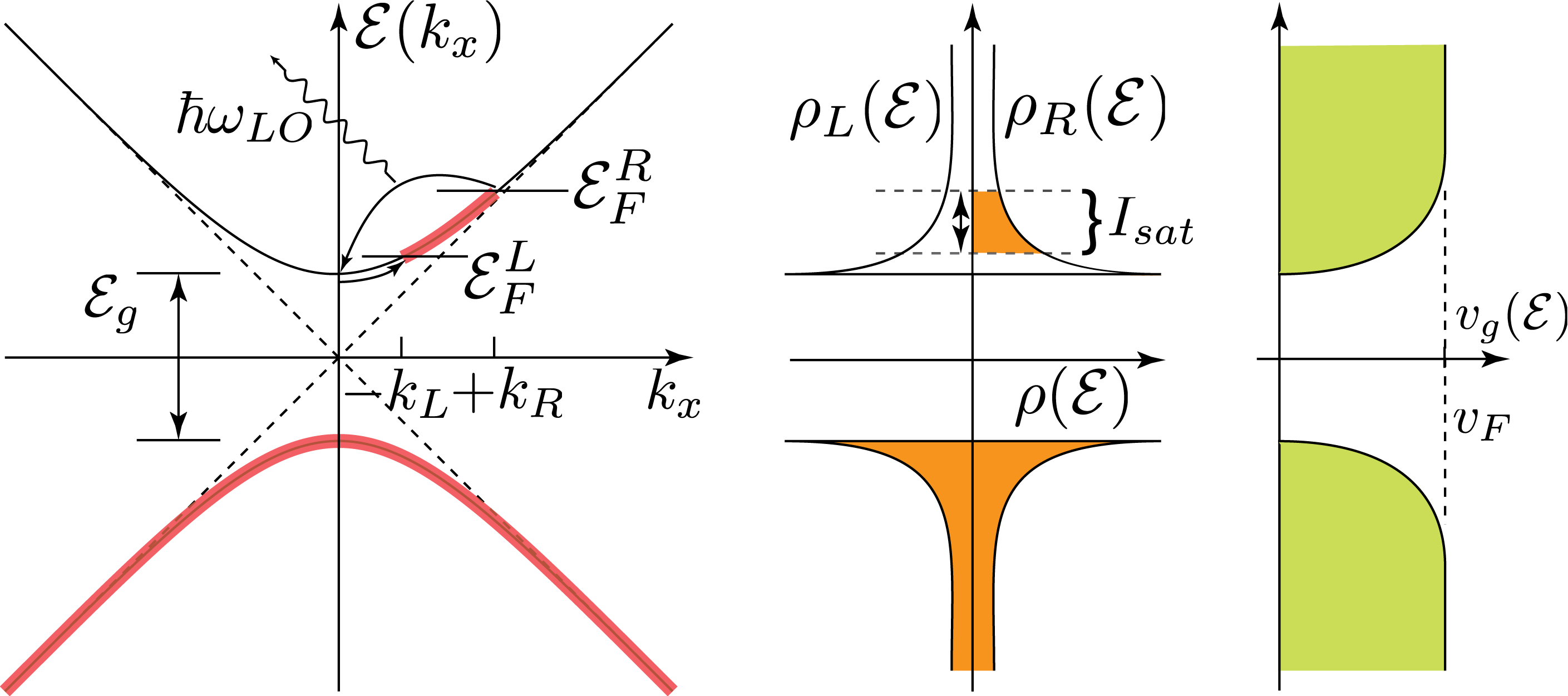}
\caption{The occupation function of carriers for a s-CNT with $n< n_{cr}$ at high fields.  The emission of optical phonons is possible when the highest occupied state and the lowest unoccupied state differ by an energy $\hbar \omega_{LO}$; the lowest unoccupied state here is the band edge.  The group velocity of carriers as a function of their energy is also shown.} 
\label{sCNT_kspace_lowdensity}
\end{center}
\end{figure}

{\em Critical parameters}: A critical Fermi level $\mathcal{E}_{F,cr}^{0}$ is the equilibrium Fermi energy such that at a high bias, $\mathcal{E}_{F}^{R} - \mathcal{E}_{F}^{L} = \hbar \omega_{LO}$, and $\mathcal{E}_{F}^{L} = \mathcal{E}_{g}/2$.  From Eq. \ref{chargebalance} the critical Fermi energy must be 
\begin{equation}
\mathcal{E}_{F,cr}^{0} = \frac{1}{2} \sqrt{ (\hbar \omega_{LO})^{2} + \hbar \omega_{LO}  \mathcal{E}_{g} + \mathcal{E}_{g}^{2} },
\label{efcrit}
\end{equation}
and is dependent only on the bandgap and the LO phonon energy of the s-CNT, which are fixed by the diameter and the lattices structure.  The corresponding equilibrium 1D critical carrier density is 
\begin{equation}
n_{cr} = \frac{g}{2 \pi \hbar v_{F}} \sqrt{\hbar \omega_{LO} ( \hbar \omega_{LO} + \mathcal{E}_{g} ) }, 
\label{ncrit}
\end{equation}
which again is a fundamental quantity for the s-CNT.  Physically, if the carrier density is above this, the LEES is a {\em left}-going state (as depicted in Figure \ref{sCNT_kspace_hidensity}), and if the carrier density is lower, then the LEES is is a {\em right}-going state, as depicted in Figure \ref{sCNT_kspace_lowdensity}.

We define two ratios: $\alpha = \mathcal{E}_{g} / 2 \mathcal{E}_{F}^{0}$ and $\beta =\hbar \omega_{LO} / 2 \mathcal{E}_{F}^{0}$.  In addition, we consider only those semiconducting tubes for which $\mathcal{E}_{g} > \hbar \omega_{LO}$ so that no interband optical phonon-assisted transitions are allowed.  This condition implies that $\beta < \alpha < 1$ when $\mathcal{E}_{F}^{0} > \mathcal{E}_{g}/2$, i.e., when the Fermi level is inside the band and carriers are available for current conduction.  If $\mathcal{E}_{F} < \mathcal{E}_{g}/2$, no current flows at $T \rightarrow 0$ K.

{\em Case 1: High carrier density}: We first consider the situation when $\mathcal{E}_{F}^{0} \geq \mathcal{E}_{F,cr}^{0}$, and $n \geq n_{cr}$.  Again, by ensuring particle number conservation and the condition $\mathcal{E}_{F}^{R} - \mathcal{E}_{F}^{L} = \hbar \omega_{LO}$, the high-bias quasi-Fermi levels for the right and left going states are found  to be 

\begin{equation}
\mathcal{E}_{F}^{R} = \mathcal{E}_{F}^{0}\sqrt{1 + \frac{ \alpha^{2} \beta^{2} }{1 -  \alpha^{2} - \beta^{2} }} + \frac{\hbar \omega_{LO}}{2},
\label{hiefr}
\end{equation}
and
\begin{equation}
\mathcal{E}_{F}^{L} = \mathcal{E}_{F}^{0}\sqrt{1 + \frac{ \alpha^{2} \beta^{2} }{1 -  \alpha^{2} - \beta^{2} }} - \frac{\hbar \omega_{LO}}{2},
\label{hiefl}
\end{equation}

{\em Case 2: Low carrier density}: If on the other hand the equilibrium Fermi level is such that $\mathcal{E}_{F}^{0} < \mathcal{E}_{F,cr}^{0}$, and $n < n_{cr}$, then the condition $\mathcal{E}_{F}^{R} - \mathcal{E}_{F}^{L} = \hbar \omega_{LO}$ cannot be satisfied due to insufficient carriers in the conduction band.  However, since the LEES is now a right-going state, the condition $\mathcal{E}_{F}^{R} - \mathcal{E}_{g}/2 = \hbar \omega_{LO}$ still allows for the emission of LO phonons.  Under this situation, the quasi-Fermi levels are given by

\begin{equation}
\mathcal{E}_{F}^{R} = \mathcal{E}_{g}/2 + \hbar \omega_{LO},
\label{loefr}
\end{equation}
and
\begin{equation}
\mathcal{E}_{F}^{L} = \mathcal{E}_{F}^{0}\sqrt{ 4 - 3 \alpha^{2} + 4 \beta^{2} + 4 \alpha \beta - 8 \sqrt{\beta} \sqrt{ 1 - \alpha^{2} } \sqrt{ \alpha - \beta } },
\label{loefl}
\end{equation}

In this condition, the HFES $\mathcal{E}_{F}^{R}$ gets pinned at the energy $\mathcal{E}_{g}/2 + \hbar \omega_{LO}$, and the LEES is at the bottom of the conduction band.  This case is depicted schematically in Figure \ref{sCNT_kspace_lowdensity}.

\begin{figure}
\begin{center}
\leavevmode 
\epsfxsize=3.4in 
\epsfbox{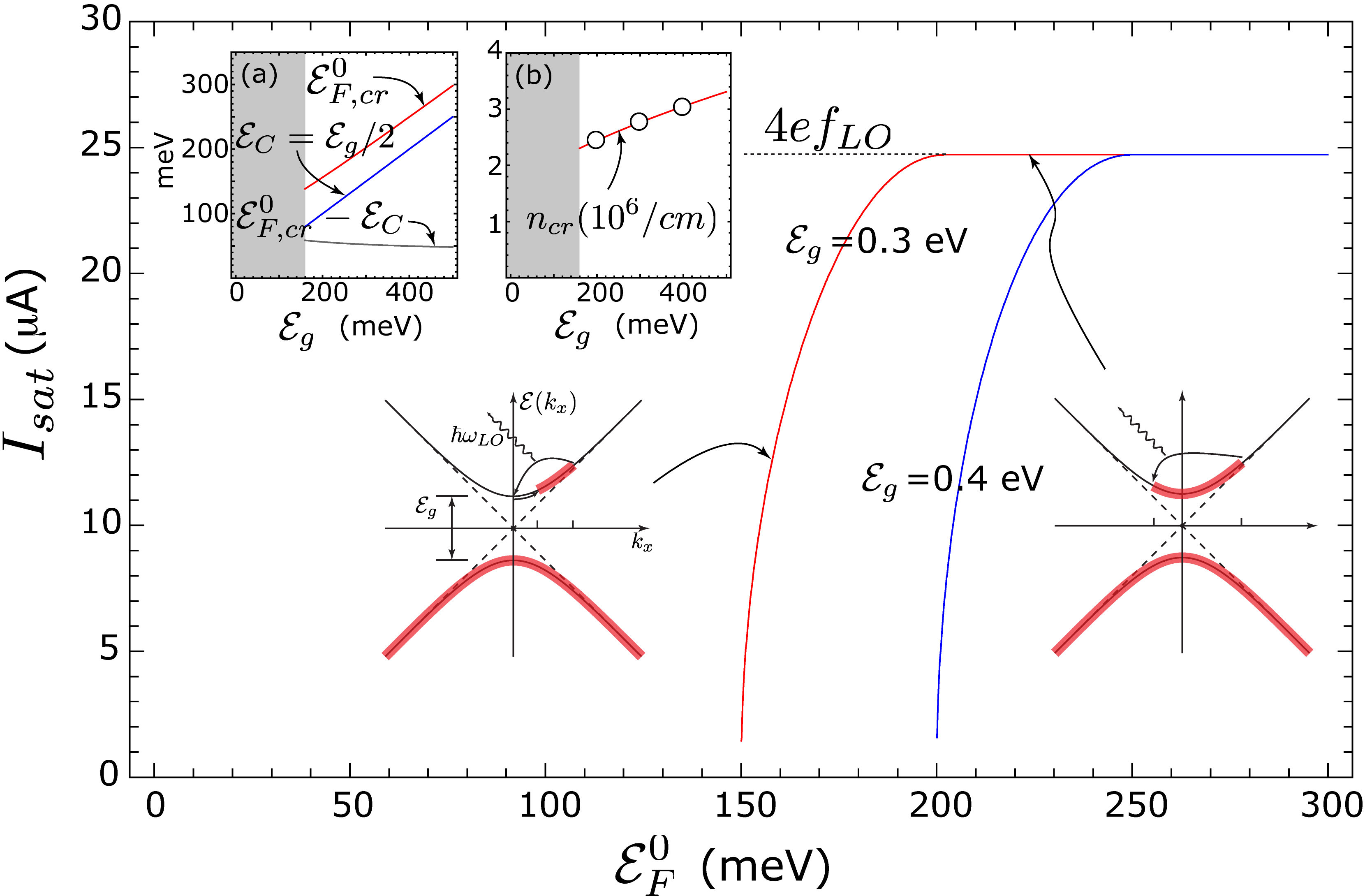}
\caption{Saturation current of s-CNTs ($\mathcal{E}_{g} = $0.3 \& 0.4 eV as a function of $\mathcal{E}_{F}^{0}$.  The insets show (a) the critical Fermi level $\mathcal{E}_{F,cr}^{0}$ and (b) the critical carrier concentration $n_{cr}$ as a function of the bandgap of the s-CNT.} 
\label{sCNT_Isat}
\end{center}
\end{figure}

{\em Saturation Current}: The current flowing through the s-CNT is given in the $T \rightarrow 0$ K limit by the relation
\begin{equation}
I_{sat} = e g \frac{1}{L} \sum_{k} f_{k} v_{k} = e \frac{g}{2 \pi} \int_{k_{L}}^{k_{R}} dk  v_{g}(k) ,
\end{equation}
where $L$ is the length of the CNT, and $v_{g}(k) = \hbar^{-1} \nabla_{k} \mathcal{E}(k) = v_{F} k_{x} / \sqrt{ k_{x}^{2} + k_{n}^{2}}$ is the projection of the group velocity  of the carriers in the direction of the electric field (Fig \ref{sCNT_kspace_lowdensity}).  The saturation current for the CNT evaluates to
\begin{equation}
I_{sat}(k_{n}, k_{F}) = e \frac{g}{ 2 \pi } v_{F} [ \sqrt{ k_{R}^{2} + k_{n}^{2}} - \sqrt{ k_{L}^{2} + k_{n}^{2}} ] ,
\end{equation}
which may be reduced the simpler form
\begin{equation}
I_{sat} = e \frac{g}{ 2 \pi } \frac{ ( \mathcal{E}_{F}^{R} - \mathcal{E}_{F}^{L} ) }{\hbar}.
\label{Isat}
\end{equation}

This is no different from what has been shown in \cite{prl00dekkerCNTsatCurrent} for m-CNTs, but it also holds for s-CNTs.  The key point here is that for low fields, corresponding to $e V_{DS} < \hbar \omega_{LO}$,  $\mathcal{E}_{F}^{R} - \mathcal{E}_{F}^{L} = e V_{DS}$, and one recovers the Landauer relation $I = g (e^{2} / h ) V_{DS}$.  However, the same relation may be used to understand saturation currents at high bias conditions as well.  Since the respective quasi-Fermi levels were shown earlier to depend on the equilibrium Fermi level (or, indirectly, the carrier density), the saturation current in a s-CNT depends on $\mathcal{E}_{F}^{0}$.  This is in stark contrast to m-CNTs, for which the quasi-Fermi level separation is $\hbar \omega_{LO}$ leading to a saturation current of $I_{sat} = 4 e f_{LO} \sim 25 \mu$A, no matter where the equilibrium Fermi level $\mathcal{E}_{F}^{0}$.

In a s-CNT, the group velocity of a fraction of carriers near the band-edge is less than or equal to $v_{F}$, whereas for a m-CNT is it {\em always} equal to $v_{F}$.  Thus, the single subband saturation current in a s-CNT can {\em never exceed} the current in a m-CNT for the same carrier density.  For high 1D carrier densities in s-CNTs (corresponding to $n > n_{cr}$ or equivalently $\mathcal{E}_{F}^{0} > \mathcal{E}_{F,cr}^{0}$ derived in Eqs \ref{efcrit}, \ref{ncrit}), the saturation current is identical to that in a m-CNT, i.e., $I_{sat} = 4 e f_{LO}$ since the condition $\mathcal{E}_{F}^{R} - \mathcal{E}_{F}^{L} = \hbar \omega_{LO}$ holds.  This similarity can be understood by noting that there is a van-Hove singularity at the band-edge of s-CNTs.  This implies that there are many more carriers that contribute to the saturation current in a s-CNT than in a m-CNT, though a fraction of those carriers move with velocities less than the Fermi velocity.  Since the DOS for a m-CNT is $\rho_{M}(\mathcal{E}) = g/ \pi \hbar v_{F}$, the number of right-going carriers over a energy bandwidth $\hbar \omega_{LO}$ is $n_{M} = 4 f_{LO} / v_{F}$.  The ratio of effective current-carrying states at saturation in a high-density s-CNT to that in a m-CNT is therefore greater than $n_{cr} / n_{M} = \sqrt{ 1 + ( \mathcal{E}_{g} / \hbar \omega_{LO} ) }$.

\begin{figure}
\begin{center}
\leavevmode \epsfxsize=3.4in \epsfbox{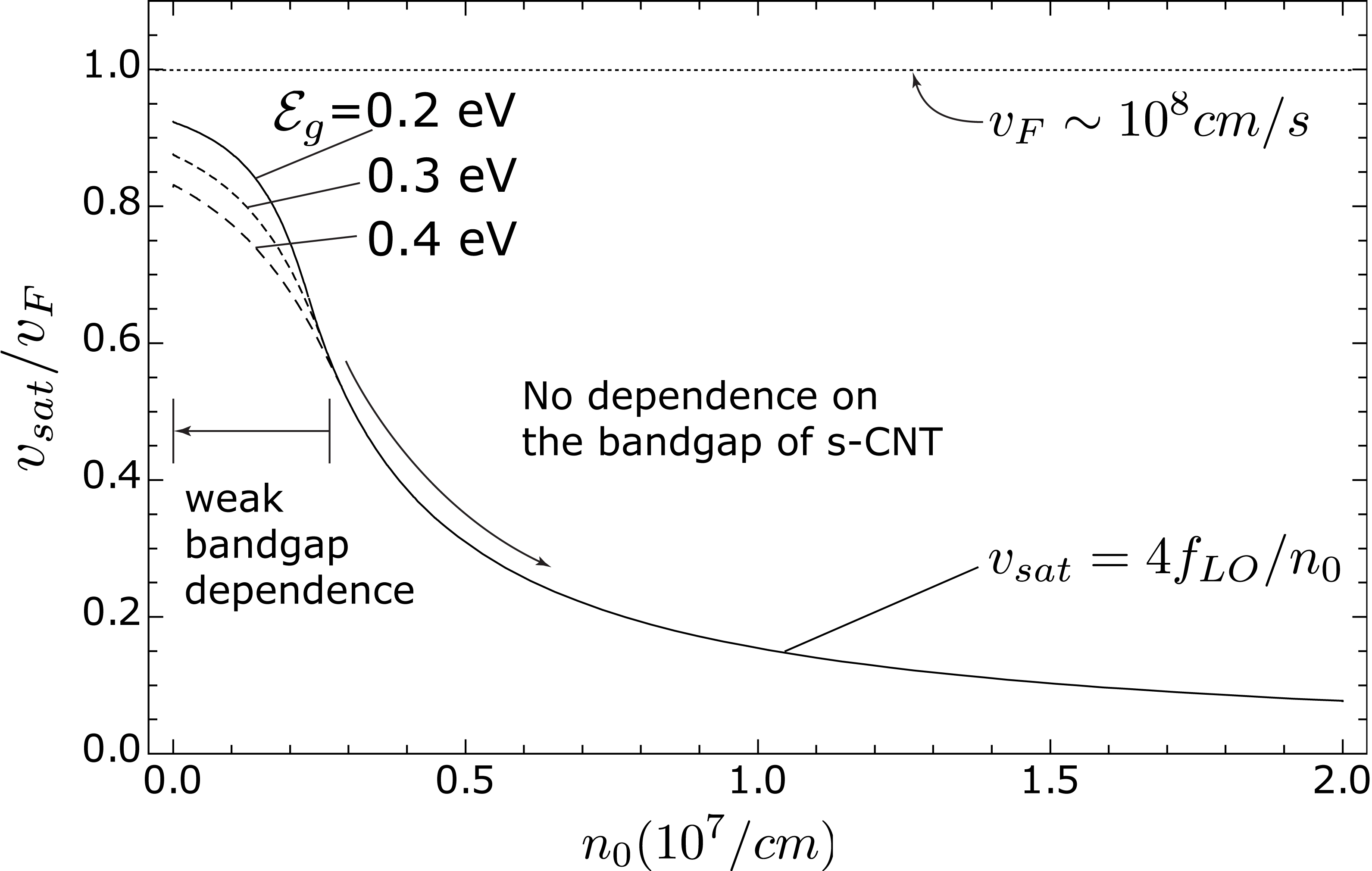}
\caption{Saturation velocity of carriers in s-CNTs with $\mathcal{E}_{g}$ = 0.2, 0.3, \& 0.4 eV as a function of the 1D carrier density.  For carrier densities $n_{0} > n_{cr}$, $v_{sat} = 4 f_{LO} / n_{0}$, and has no dependence on the bandgap of the s-CNT; for $n_{0} < n_{cr}$, the saturation velocity has a weak dependence on $\mathcal{E}_{g}$.} 
\label{sCNT_vsat_n1d}
\end{center}
\end{figure}

For carrier densities in s-CNTs less than $n_{cr}$, the saturation current is given by Eq. \ref{Isat}, with $\mathcal{E}_{F}^{R}, \mathcal{E}_{F}^{L}$ from Eqs. \ref{loefr} \& \ref{loefl}.  This dependence exemplifies how a s-CNT under a high source-drain bias `switches-off' when the Fermi level is pulled down towards the band-edge and then into the gap, for example, by electrostatic gating.  The dependence of the saturation current on $\mathcal{E}_{F}^{0}$ is plotted in Fig \ref{sCNT_Isat} for s-CNTs for two representative bandgaps.  The insets show the critical Fermi level and the critical 1D carrier concentration above which the saturation properties of the s-CNT become indistinguishable from that of a m-CNT.

\begin{figure}
\begin{center}
\leavevmode \epsfxsize=3.4in \epsfbox{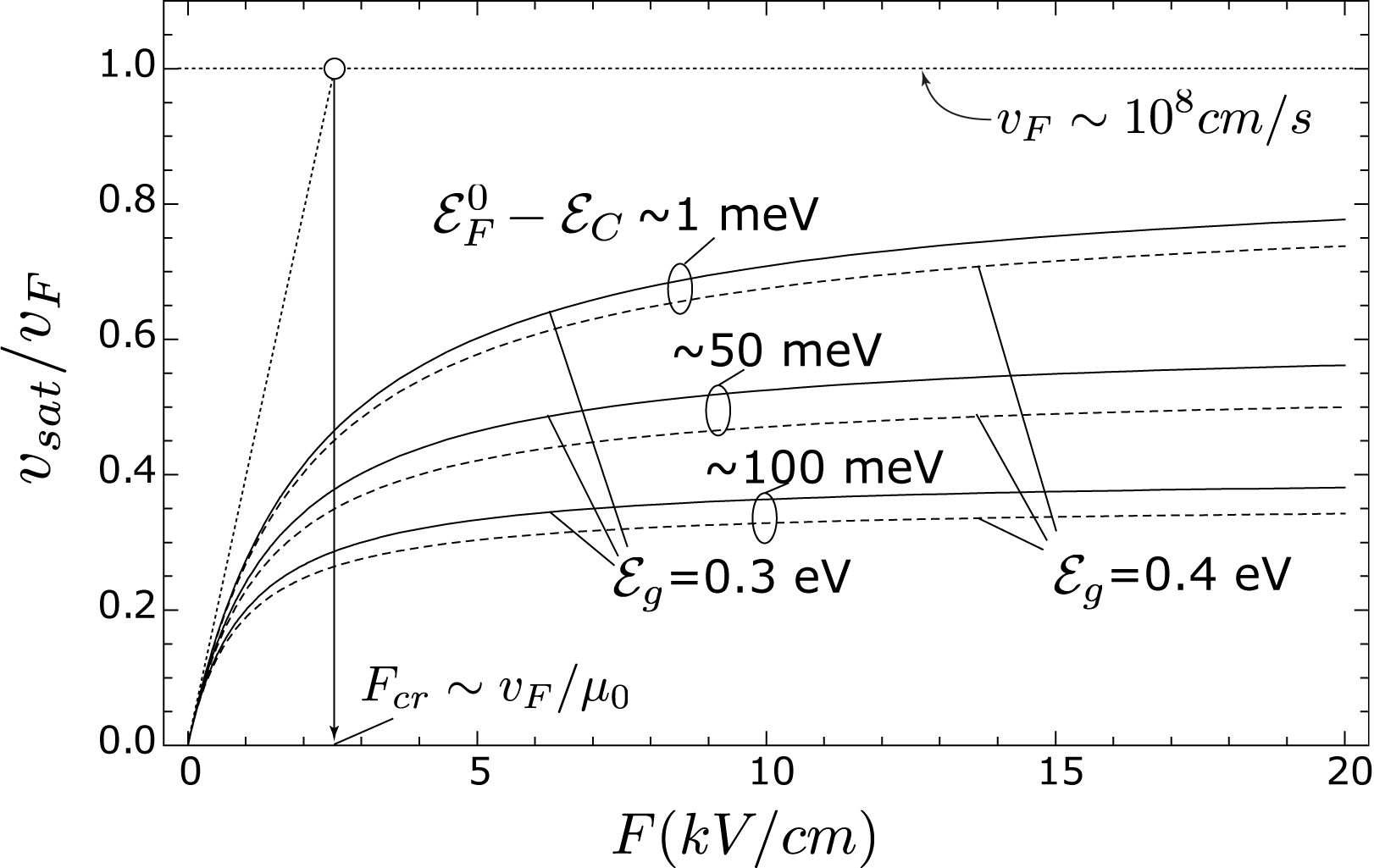}
\caption{The velocity-field curves for carriers in s-CNTs of two bandgaps (solid line: $\mathcal{E}_{g} = 0.3$ eV, dashed line: $\mathcal{E}_{g} = 0.4$ eV.  The strong dependence of the velocity saturation on the effective carrier concentration, as measured by $\mathcal{E}_{F}^{0} - \mathcal{E}_{C}$ is indicated.  The low-field mobility assumed is $\mu_{0} = 40,000$ cm$^{2}$/V.s for the calculation.} 
\label{sCNT_v_F}
\end{center}
\end{figure}
{\em Saturation velocity and velocity-field curves:} Under the assumption that all carriers in the conducting band contribute to the saturation current equally, the saturation velocity of carriers may be written as $v_{sat} = I_{sat} / e n_{0}$, which evaluates to
\begin{equation}
v_{sat} = v_{F} \cdot \frac{ \frac{1}{2} (\mathcal{E}_{F}^{R} -  \mathcal{E}_{F}^{L} ) }{\sqrt{ (\mathcal{E}_{F}^{0})^{2} - (\frac{\mathcal{E}_{g}}{2})^{2} }} = \frac{g}{2 \pi \hbar n_{0}} \cdot (\mathcal{E}_{F}^{R} -  \mathcal{E}_{F}^{L} ),
\label{vsat}
\end{equation}
and is strongly dependent on the 1D carrier density in the band.  Fig \ref{sCNT_vsat_n1d} shows the dependence of the saturation velocity on the density as well as the corresponding equilibrium Fermi level for s-CNTs of varying bandgaps.  For carrier densities $n<n_{cr}$, it increases and approaches $v_{F}$ as $n \rightarrow 0$ and $\mathcal{E}_{g} $ decreases.  However, from Eq. \ref{vsat}, the saturation velocity at carrier densities $n > n_{cr}$ is given by $v_{sat} = 4 f_{LO} / n_{0}$, and is inversely proportional to the 1D carrier density.  The theory is able to explain why the saturation velocity becomes {\em independent} of the bandgap of the s-CNT, as was found in numerical simulations of the BTE in \cite{prl05avourisCNTsatCurrent}.

If the low-field carrier mobility is $\mu_{0}$, then a critical electric field may be defined as $F_{cr} = v_{sat} / \mu$; the net velocity-field curve may then be approximated by the relation 
\begin{equation}
v = \frac{ v_{sat} }{ 1 + (v_{sat} / \mu_{0} F)  },
\end{equation}
which is strongly dependent on the carrier density and the bandgap of the s-CNT.  Fig \ref{sCNT_v_F} shows a few velocity-field curves plotted for s-CNTs of bandgaps 0.3 and 0.4 eV, for different carrier densities, as indicated by the energy $\mathcal{E}_{F}^{0} - \mathcal{E}_{C}$.  The low-field electron mobility chosen for this calculation is $\mu_{0} = 40,000$ cm$^{2}$/V.s, which is determined primarily by elastic scattering processes (acoustic phonons \& impurities).  It is emphasized that the often-used concept of saturation velocity is an {\em ensemble property}, and care must be exercised in using it to predict carrier transit times, switching speeds, and allied quantities since there are carriers that move at velocities higher as well as lower than this velocity.  

{\em Deviations from the theory:}  Within the limits of the theory presented, the saturation current in a s-CNT due to single subband conduction in the diffusive limit can never exceed $I_{0} = 4 e f_{LO}$.  Any experimental observation of higher saturation currents in s-CNTs {\em must} be therefore attributed to effects that have not been considered.  Five possible factors are listed: a) non-uniform heating due to ultrafast LO phonon emission \cite{prl05leburtonCNTsatCurrent} and/or hot-phonon effects \cite{prl05popHotPhCNT}, b) ballistic effects (i.e., if the length of the CNT is shorter than the mean-free path for LO-phonon emission $L < \mathcal{L}_{LO}$) \cite{nl04mcEuenCNTphononBallistic}, c) occupation of multiple subbands, d) contribution from band-to-band Zener tunneling \cite{prl04bachtoldCNTsatCurrent} and/or impact ionization, and e) space-charge injection into the s-CNT by the contacts.  These factors can be incorporated into the analytical framework presented here, but is not considered in this letter.

{\em Conclusions:} In conclusion, a simple analytical theory is presented that shows that current saturation in s-CNTs and m-CNTs are indistinguishable if the carrier density in the s-CNT is above a critical value determined by the bandgap and the LO phonon energy of the s-CNT.  The ensemble saturation velocity is found to be independent of the bandgap of the s-CNT for such carrier densities as well.  The author acknowledges support from the NSF CAREER award (DMR-0645698), and useful discussions with Dr. Chagaan Bataar (ONR).

\bibliographystyle{unsrt}


\pagebreak

\end{document}